\documentclass[english]{article}
\usepackage{mathptmx}
\usepackage[T1]{fontenc}
\usepackage[latin9]{inputenc}
\usepackage{amsthm}
\usepackage{amsmath}

\makeatletter
\usepackage{hyperref}
\usepackage[margin=1in]{geometry}
\numberwithin{equation}{section}
\usepackage{graphicx}	

\makeatother

\usepackage{babel}

\begin{document}

\title{Ultraviolet-Finite QFT on Curved Space-Times}

\author{Philip Tillman \\
 email: phil.tillman@gmail.com}

\date{\today }
\maketitle
\begin{abstract}
This paper proposes a different interpretation on the renormalization
program of an interacting Klein-Gordon field in curved space-times.
Rather than being just another renormalization program we argue that
it that it makes the QFT ultraviolet-finite at its foundation. More
precisely, we argue that renormalization is a part of the foundation
of QFT and that without it the theory is mathematically ill-defined
or at the very least incompletely specified.
\end{abstract}

\section{Introduction}
\begin{quotation}
\begin{center}
\emph{{}``Most physicists are very satisfied with the situation.
They say: 'Quantum electrodynamics is a good theory, and we do not
have to worry about it any more.' I must say that I am very dissatisfied
with the situation, because this so-called 'good theory' does involve
neglecting infinities which appear in its equations, neglecting them
in an arbitrary way. This is just not sensible mathematics. Sensible
mathematics involves neglecting a quantity when it is small -- not
neglecting it just because it is infinitely great and you do not want
it!'' Dirac, source: {}``Dirac'' By Helge Kragh 1990}
\par\end{center}
\end{quotation}
In renormalization we subtract infinite constants and absorb infinite
constants into the  parameters. Any person who first sees this method
is somewhat unsettled. Surely it works, because the predictions back
it up, but we want to understand exactly why it works. In any physical
theory we cannot simply {}``throw out'' the infinities because we
don't like them.

The presence of infinities in it of itself is not a reason to disallow
them because there is a reason that they are there. For instance,
in GR we have singularity at the event horizon of a black hole solution
provided by Schwarzschild. After much debate it is now accepted that
this a defect of the chosen coordinates. The Kruskal extension of
Schwarzschild coordinates is coordinate system that has no infinities
at the event horizon. The event horizon, as we now know, is simply
a region of space after which nothing can escape, not a singularity.
However there is a more serious singularity (the point of infinity
density) at the center of the black hole, but it is assumed that another
theory, namely a unified theory or quantum gravity, takes hold at
such densities and energies. Thus, infinities usually represent poor
choices or limits of the theory and in the latter case their very
presence means that some other theory must take over to properly describe
the physics of these regions.

Renormalization is very different. When we try compute things that
we should be able to compute, such as mass, they blow up to infinity.
The only way, at the time it was formulated, to save the theory was
to concoct a semi-mathematical method for absorbing or subtracting
infinities. Some of the founders of QFT, such as Dirac, were very
unsettled by these turn of events. In mathematics you cannot absorb
or subtract infinities in this way. Despite all this renormalization
works fantastically well when it comes to predictions in experiments.
Thus, it is a consensus that renormalization could be (in the future)
be put on firm mathematical ground and all can be justified, even
while there maybe some iffy mathematics now in renormalization. This
is the current viewpoint of most physicists today.

In Minkowski space QFT it is the interpretation taken by Scharf {[}\hyperlink{Scharf}{2}{]}
that the Epstein-Glaser perturbative renormalization {[}\hyperlink{EpsteinGlaser}{1}{]}
is not merely another renormalization program -- it makes QFT ultraviolet-finite.
By improperly incorporating causality and splitting distributions
into retarded and advanced parts, such as propagators, incorrectly
you are punished by ultraviolet infinities. This program has been
extended to curved space-time in {[}\hyperlink{math-ph/9903028}{5}{]}.
To accomplish this two crucial requirements are needed: one is the
proper inclusion of causality and the other is the micro-local spectrum
condition {[}\hyperlink{math-ph/9903028}{5}{]}.%
\footnote{The relation to Scharf's ideas are seen by the connection between
the splitting of distributions (e.g. propagators) and the micro-local
spectrum condition. The micro-local spectrum can be thought of as
requiring that every propagator appearing in Feynman diagrams must
be of Hadamard form.%
} It is the author's opinion that Scharf's viewpoint holds in this
case as well as we will argue.

Infinities in renormalization come two basic types: infrared and ultraviolet.
The former regards more to questions on states such as the existence
of vacuum states, thermal states, etc. (e.g. see {[}\hyperlink{Buchholz}{6}{]}).
The latter is a more serious problem from a foundational perspective,
but in {[}\hyperlink{math-ph/9903028}{5},\hyperlink{gr-qc/0103074}{3},\hyperlink{gr-qc/0404074}{4}{]}
a new paradigm for perturbative renormalization of interacting of
QFT in curved space-times (QFTCS) has been defined to eliminate the
ultraviolet infinities. The main insight is that by taking improper
pointwise products of distributions you are punished by ultraviolet
infinities. For example, the pointwise product of the Feynman propagator
is \emph{never} well-defined and so you would get ultraviolet infinities
in that case. It is this new program that this paper focuses on.

With regards with renormalization, we have made major gains in understanding
the sources of the infinities and where the current standard Minkowski
space techniques go wrong. With respect to ultraviolet infinities,
we could view this new correction of the infinities as another renormalization
scheme of a QFTCS. However, the author proposes a different interpretation
that based on the observation that the ultraviolet infinities arise
from ill-defined products of distributions. Thus the {}``unrenormalized''
theory is objectionable because these products are not well-defined
in a mathematical sense. Thus, we could view the original {}``unrenormalized''
theory as being not a viable candidate for a physical theory on these
grounds. This then suggest that we should incorporate the definitions
of products of distributions into the foundations of the theory. QFT
becomes ultraviolet-finite since the renormalization program is {}``cooked''
into it a priori.

In the sense above, the term renormalization is a misnomer. By saying
this, the author does not advocate to get rid of the term -- it's
here to stay whether we like it or not. What the author simply wants
is to rethink about the term itself and what previously meant and
what it should mean now. Previously it meant that we start with the
{}``unrenormalized theory'' that is (as we believe it to be) correct
as a physical theory, compute infinities everywhere, and get rid of
them through renormalization to obtain the correct theory we actually
use. In this new paradigm the {}``unrenormalized theory'' cannot
be regarded as even a candidate for a physical theory -- because at
the very least it is not specified well enough. Therefore, the author
argues that renormalization is not only needed to obtain usable predictions,
but justified both mathematically and physically in the specification
of QFT. Renormalization should be viewed as a program to properly
define the observable algebra in QFT.

To say that renormalization is misnomer is not to say that there are
not any regularizations involved, but simply without it (including
any regularization involved) QFT on curved space-time is simply mathematically
incorrect or at the very least incomplete. The way to think of this
interpretation is that it is a validation of what physicists already
suspected: renormalization is correct and can be put on a solid mathematical
foundation. This new interpretation of the renormalization program
of QFTCS allows us to put it on a solid mathematical foundation and
into the very definition of QFT.

\section{An Interpretation of Perturbative Renormalization of QFTCS: Ultraviolet-Finite
Perturbative QFT}

In this section we define $\mathcal{W}\left(M,g\right)$ as the algebra
of Wick polynomials, their time-order products, and derivatives constructed
in {[}\hyperlink{gr-qc/0103074}{3},\hyperlink{gr-qc/0404074}{4}{]}.
We denote the constructions of the Wick powers and their time-ordered
products presented in these references by the maps $N^{k}\left[\varphi\right]=\varphi^{k}$
and $T\left(\varphi^{k_{1}}\cdots\varphi^{k_{n}}\right)$ respectively.

The maps $N^{k}$ and $T$, the author claims, are fundamental to
the specification of QFT. A possible issue arises because these maps
have ambiguities in their definition, but this will not cause problems
if the physical theory is independent of the choice. If this were
the case then the physical predictions would be affected by the choice
of maps $N^{k}$ and $T$. Similar to choice of coordinates, physical
theories must be independent of such artificial choices and we argue
that this is the case here. The reasoning is based on the observation
in {[}\hyperlink{gr-qc/0103074}{3}{]} that the ambiguities of $N^{k}$
and $T$ are precisely the renormalization ambiguities.

Specifically they say the following. Let $\varphi^{k}$, $T\left(\prod_{i}\varphi^{k_{i}}\right)$
and $N_{H}^{k}\left[\varphi\right]=:\varphi^{k}:{}_{H}$, $\tilde{T}\left(\prod_{i}:\varphi^{k_{i}}:_{H}\right)$
be two different prescriptions for defining Wick powers and their
time-ordered products. Let the S-matrix of our theory be\[
S\left(\mathcal{L}_{int}\right)=1+\sum_{n\geq1}\frac{i^{n}}{n!}\int_{M^{n}}T\left(\mathcal{L}_{int}\left(x_{1}\right)\cdots\mathcal{L}_{int}\left(x_{n}\right)\mu_{g}\left(x_{1}\right)\cdots\mu_{g}\left(x_{n}\right)\right)\]
 where interaction part of the Lagrangian is $\mathcal{L}_{int}\left(x\right)=f\varphi^{4}\left(x\right)$
and the S-matrix in the other prescription is \[
\tilde{S}\left(\tilde{\mathcal{L}}_{int}\right)=1+\sum_{n\geq1}\frac{i^{n}}{n!}\int_{M^{n}}\tilde{T}\left(\tilde{\mathcal{L}}_{int}\left(x_{1}\right)\cdots\tilde{\mathcal{L}}_{int}\left(x_{n}\right)\mu_{g}\left(x_{1}\right)\cdots\mu_{g}\left(x_{n}\right)\right)\]
where $\tilde{\mathcal{L}}_{int}\left(x\right)=f:\varphi^{4}\left(x\right):_{H}$.
Note that $f$ is a coupling function that is constant in some compact
region of space (our interaction center).%
\footnote{This is an assumption of the renormalization program which allows
prevent infrared infinities by truncating the S-matrix series. The
limit of coupling functions to constants is the adiabatic limit and
in this limit the infrared infinities must be dealt with.%
}

In {[}\hyperlink{gr-qc/0103074}{3}{]} they showed that the ambiguities
of the definitions of the Wick powers and their time-ordered products
are precisely the renormalization ambiguities. I.e., that $\tilde{S}\left(\tilde{\mathcal{L}}_{int}\right)=S\left(\mathcal{L}_{int}+\delta\mathcal{L}_{int}\right)$
for some local and covariant field $\delta\mathcal{L}_{int}$ which
has the same form as the original Lagrangian $\mathcal{L}=\mathcal{L}_{0}+\mathcal{L}_{int}$.
Thus the theories are equivalent and are related by a redefinition
of the constants like the field strength and couplings (apart from
terms proportional to the identity which contribute to the overall
phase of $S$ which are inconsequential for QFT). Thus, since the
constants must be measured then the physical predictions will always
be the same regardless of the choice of $N^{k}$ and $T$. It is observed
that:
\begin{itemize}
\item \emph{A physical model which is renormalizable (i.e., their Lagrangians
are of renormalizable form), such as Klein-Gordon with a $\varphi^{4}$
interaction term, together with all other axioms specified in }{[}\hyperlink{gr-qc/0103074}{3},\hyperlink{gr-qc/0404074}{4}{]}\emph{
uniquely define the physical theory.}
\end{itemize}
Subsequently, suppose that we assume that renormalization is required
for the specification of QFT\emph{ then} \emph{only theories which
are renormalizable can be candidates for the physical theories}. This
should be viewed as starting point for QFT so that we disallow non-renormalizable
theories and, therefore, inconsistent physical predictions.

\medskip{}

The idea of ultraviolet-finite QFT is as follows: 

Without any proper prescription for both $N^{k}$ and $T$ we do not
have a mathematically well-defined theory, thus they are essential.
This means that the {}``unrenormalized theory'' cannot be even a
possible candidate for a physical theory. As a result of ill-defined
products of distributions we would be punished by ultraviolet infinities. 

If we assume that the maps $N^{k}$ and $T$ are in the foundations
of the theory then ambiguities are no problem only for Lagrangians
of renormalizable form ($\delta\mathcal{L}_{int}$ which has the same
form as the original Lagrangian). Then no matter what prescription
you choose, say $\varphi^{k}$, $T\left(\prod_{i}\varphi^{k_{i}}\right)$,
the theory will be physically equivalent to all other choices since
the form of the Lagrangian (and hence the S-matrix) is the same. Only
the constants differ, but these are always fixed by experiments. In
this way, the physics is completely determined regardless of prescription
we choose. The {}``renormalized theory'', and not the {}``unrenormalized
theory'', can be the only candidate for a physical theory by which
we can make predictions.

We now can say the label renormalization is a misnomer, since infinities
never arise when we compute any quantities that should be finite like
mass or the coupling. This is seen by the fact that the renormalization
scheme $N^{k}$ and $T$ is not an afterthought. We don't compute
quantities, such as mass, in what we assume is a perfectly well-defined
theory, suddenly get infinities, and then have to rid ourselves of
them somehow. The renormalization step is cooked into the very definition
of the theory. \emph{Renormalization allows us to define} $\mathcal{W}$
which is the observable algebra. The ambiguities (which are finite)
are not a problem (in terms of physical predictions) if we assume
that the S-matrix is renormalizable since the form of the Lagrangian
is the same. It is then only a matter of redefining our constants
and then going out and use experiments to fix them.

\section{Summary}

The results of this paper does take an unconventional viewpoint, but
the insights are based on quite conventional results of QFT on curved
space-time. Namely that QFT is at the very least incompletely specified
without renormalization, because operations with distributions (like
products) must be carefully defined. That is renormalization is not
only now mathematically well-defined, but it is {}``cooked'' into
the foundations of QFT -- i.e. it is part of QFT's definition. The
author suggests that we should require that all candidates for physical
theories to be ones which are renormalizable (i.e., their Lagrangians
are of renormalizable form). This is to say that if a QFT makes physical
and mathematical sense it must be renormalizable, and this should
be a starting principle of QFT.

If renormalization is a part of the definition of QFT then there are
no ultraviolet infinities. This is not to say that there are no regularizations
in the defining the Wick powers and their time-ordered products, there
clearly are. These regularization are not as a result of computing
quantities that should be finite, but are not. Quite the contrary,
it is the definition of the framework, i.e. the algebra of observables
$\mathcal{W}$, that only uses these regularizations. So what the
author proposes is that we interpret these regularizations as not
only needed, but justified (both mathematically and physically) to
be able to formulate QFT correctly. Having renormalization built into
the theory becomes ultraviolet-finite.

\section{References}

\noindent{[}\hypertarget{EpsteinGlaser}{1}{]} H. Epstein, and V.
Glaser:Adiabatic limit in perturbation theory. In: G. Velo and A.
S. Wightman, (eds.) Renormalization Theory. Proceedings, Dordrecht-Holland:
D. Reidel Publishing Co., 1976

\noindent{[}\hypertarget{Scharf}{2}{]} G.Scharf, Finite Quantum
Electrodynamics, Texts and Monographs in Physics, Springer-Verlag
1989

\noindent{[}\hypertarget{gr-qc/0103074}{3}{]} S. Hollands and R.
M. Wald: \textquotedblleft{}Local Wick Polynomials and Time Ordered
Products of Quantum Fields in Curved Space,\textquotedblright{} Commun.
Math. Phys. 223, 289-326 (2001), \href{http://arxiv.org/abs/gr-qc/0103074}{gr-qc/0103074},
S. Hollands and R. M. Wald: \textquotedblleft{}Existence of local
covariant time-ordered-products of quantum fi{}elds in curved spacetime,\textquotedblright{}
Commun. Math. Phys. 231, 309-345 (2002), \href{http://arxiv.org/abs/gr-qc/0111108}{gr-qc/0111108}

\noindent{[}\hypertarget{gr-qc/0404074}{4}{]} S. Hollands and R.
M. Wald, \textquotedblleft{}Conservation of the stress tensor in interacting
quantum fi{}eld theory in curved spacetimes,\textquotedblright{} Rev.
Math. Phys. 17, 227 (2005) \href{http://arxiv.org/abs/gr-qc/0404074}{gr-qc/0404074}

\noindent{[}\hypertarget{math-ph/9903028}{5}{]} R. Brunetti, K.
Fredenhagen: \textquotedblleft{}Microlocal Analysis and Interacting
Quantum Field Theories: Renormalization on physical backgrounds,\textquotedblright{}
Commun. Math. Phys. 208, 623-661 (2000), \href{http://arxiv.org/abs/math-ph/9903028}{math-ph/9903028}

\noindent{[}\hypertarget{Buchholz}{6}{]} D. Buchholz: The physical
state space of quantum electrodynamics, Commun. Math. Phys. \textbf{85}
(1982) 49)
\end{document}